\documentclass[12pt,preprint]{aastex}

\newcommand{\ndhm}{N$_2$H$^+$}
\newcommand{\methanol}{CH$_3$OH}
\newcommand{\ammonia}{NH$_{3}$}
\newcommand{\hcom}{HCO$^+$}
\newcommand{\cthd}{C$_3$H$_2$}
\newcommand{\cdo}{C$^{18}$O}
\newcommand{\htcom}{H$^{13}$CO$^+$}

\newcommand{\dmu}{GEVWH}

\newcommand{\kms}{km~s$^{-1}$}

\newcommand{\rin}{$R_\mathrm{in}$}
\newcommand{\rout}{$R_\mathrm{out}$}
\newcommand{\vinf}{$V_\mathrm{inf}$}

\shorttitle{The  dense core ahead of HH 80N}
\shortauthors{Masqu\'e et al.}

\begin{document}

\title{The molecular emission of the irradiated star forming core ahead of
HH 80N}

\author{Josep M. Masqu\'e\altaffilmark{1}, Josep M. Girart\altaffilmark{2}, Maria T.
Beltr\'an\altaffilmark{1}, Robert Estalella\altaffilmark{1} \and Serena
Viti\altaffilmark{3}}

\altaffiltext{1}{Departament d'Astronomia i Meteorologia, Universitat de Barcelona, 
Mart\'i i Franqu\`es 1, 08028 Barcelona, Catalunya, Spain}

\altaffiltext{2}{Institut de Ci\`encies de l'Espai, (CSIC-IEEC), 
Campus UAB, Facultat de Ci\`encies, Torre C5 - parell 2, 
08193 Bellaterra, Catalunya, Spain}

\altaffiltext{3}{Department of Physics and Astronomy, University College London,
London, WC1E 6BT, UK}

\begin{abstract}

We present a Berkeley-Illinois-Maryland Association (BIMA) Array molecular survey of
the star forming core ahead of HH 80N, the optically obscured northern counterpart
of the Herbig-Haro objects HH 80/81. Continuum emission at 1.4~mm and 8~$\mu$m is
detected at the center of the core, which confirms the presence of an embedded very
young stellar object in the core.  All detected molecular species arise in a
ring-like structure, which is most clearly traced by CS~(2--1) emission. This
molecular ring suggests that strong molecular depletion occurs in the
inner part of the core (at a radius of $\simeq 0.1$~pc and densities higher
than $\sim$~5~$\times$~$10^4$~cm$^{-3}$). Despite of the overall morphology and
kinematic similarity between the different species, there is significant molecular
differentiation along the ring-like structure. The analysis of the chemistry along
the core shows that part of this differentiation may be caused by the UV
irradiation of the nearby HH 80N object, that illuminates the part of the core
facing HH 80N, which results in an abundance enhancement of some of the
detected species.

\end{abstract}


\section{Introduction}

IRAS $18162-2048$ is a high mass protostar, with a luminosity of 
$\sim~2~\times~10^4$~$L_\odot$, powering the most luminous and largest Herbig-Haro
(HH) system,  HH~80/81/80N, associated with a very highly collimated radio jet
(Rodr\'iguez et al.\ 1980; Reipurth \& Graham\ 1988; Mart\'{i} et al.\ 1993). This
source is located in the GGD 27 region,  at a distance of 1.7 kpc in Sagittarius
(Rodr\'iguez et al.\ 1980). A far-IR spectroscopic study of the HH 80/81/80N
system shows that the FUV field radiated by the ionized material in the shock
recombination region is able to induce the formation of a photodissociation region
(PDR) in the surrounding medium of the HH objects and the jet flow (Molinari et
al.\ 2001). The HH 80N object has not been detected in the optical neither at
near infrared wavelengths, which is likely a consequence of the
extinction due to a molecular cloud in the foreground (Mart\'i et al.\ 1993). 

A dense clump of 20~$M_\odot$ was found ahead of HH~80N, firstly detected in ammonia
(Girart et al.\ 1994) and afterward detected in other species, such as CS and HCO$^+$
(Girart et al.\ 1998). BIMA array observations (with 10\arcsec\ of angular resolution)
carried out by Girart et al.\ (2001) (hereafter \dmu) showed that the CS~(2--1)
emission traces a ring-like structure, with a radius of 0.24~pc, seen edge on. This
observed morphology is most likely  the result of a strong CS depletion in the inner
region of the core (\dmu). The analysis of the CS kinematics suggests that this
structure is contracting with an infall velocity of 0.6~\kms. Assuming a gas
temperature of 17~K, equivalent to the ammonia rotational temperature (Girart et al.\
1994) the sound speed is $\sim$~0.3~\kms, implying supersonic collapse. In addition,
the CO (1--0) emission reveals a bipolar outflow that implies the presence of an
embedded protostellar object within the HH 80N core (\dmu).


In recent years, observations have found that a number of HH objects have
associated molecular condensations. Typically, these clumps are cool
($\sim$~10--20~K), dense ($\ga 3\times10^4$~cm$^{-3}$), small (sizes of $\la$~0.1
pc and masses of $\la$~1~$M_\odot$), starless and show little or no evidence of
dynamical interaction with the stellar jet (e.g. HH 1/2: Torrelles et al.\ 1994;
Viti et al.\ 2006;  HH 34: Anglada et al.\ 1995). They are possibly of the same
type as the transient small clumps found by Morata et al.\ (2005) in L673. The high
molecular abundances of some species (e.g., \methanol, \ammonia, \hcom) found in
these clumps (Girart et al.\ 2005; Viti et al.\ 2006) suggest a chemical alteration
induced by the UV radiation incoming from the HH object (Taylor \& Williams\ 1996;
Viti \& Williams\ 1999; Viti et al.\ 2003). However, BIMA observations towards the
HH 2 region reveal that despite the generally quiescent nature of the molecular
condensations ahead of the HH objects, complex dynamical and radiative interactions
occur in this region (Girart et al.\ 2005). The dense core ahead of HH 80N seems to
be one of these examples of irradiated cores (Girart et al.\ 1994, 1998), but it
has two peculiarities that make this region an interesting target to study. First,
the HH 80N core is significantly larger and more massive than those described
above, and shows star formation signatures (\dmu). This suggests a more complex
scenario involving a variety of physical processes. Second, the core shows
supersonic infall velocity, which differs from what standard contracting core
models predict (e.g. Shu et al.\ 1987; Basu \& Mouschovias\ 1994; Ciolek \&
Mouschovias\ 1995) or what is observed in other contracting cores (Williams et al.\
1999; Belloche et al.\ 2002; Furuya et al.\ 2006; Swift et al.\ 2006). This raises
the question whether the outflow has triggered or at least sped up the collapse of
the core (GEVWH).  

In this work we present the study of the emission of different molecular species
observed in the HH 80N core with the BIMA interferometer. Our aim is to
characterize the properties of the emission for different molecular tracers and to
establish their deviations from the CS~(2--1) emission analyzed in \dmu. We also
attempt to provide some possible explanations for the origin of these observed
differences by means of a qualitative comparison with other regions, but with
special attention to the HH 80N influence. A subsequent analysis of the physical
properties as well as a modeling of the chemistry of the core ahead of HH 80N will
be reported in a forthcoming paper. In \S~2 we present our observations. Their
analysis is shown in \S~3. In \S~4 we discuss the results. Finally, in \S~5, we
summarize our conclusions.

\section{Observations}

The 10-antenna BIMA array observations were carried out between November 1999 and
May 2001. The phase calibrator used was the QSO J1733$-$130 and the flux calibrators
were Mars, Uranus or Neptune depending on the epoch. A total of 12 frequency
setups were used, one at 218.2 GHz and the rest covering frequencies between 85.1
and 109.8 GHz. All the frequency setups were observed in the C configuration apart
from the 93.1~GHz frequency setup which was observed in the D configuration. The C
configuration had baselines between 6.3 to 100~m, providing  an angular resolution
of $\sim$~5$\arcsec$ and $\sim$~9$\arcsec$ at 1~mm and 3~mm, respectively. The D
configuration had baselines between 6.3 and 30~m, providing an angular  resolution
of $\sim$~25$\arcsec$ at 3~mm. Each frequency setup has upper and lower frequency
bands. The digital correlator was configured to sample part of the 800~MHz wide IF
passband in several windows. The typical window for the line observation was
configured with a 25 MHz bandwidth and 256 channels, giving a spectral resolution
typically of $\sim$~100~kHz, which corresponds to a velocity resolution of
$\sim0.15$~\kms\ and $\sim0.3$~\kms\ observing at 1~mm and 3~mm, respectively. For
the continuum maps, typical windows of 0.1 GHz divided in 32 channels were used.

The phase tracking center of the observations was set at 
$\alpha(J2000)=18^\mathrm{h}19^\mathrm{m}18\fs618$ and
$\delta(J2000)=-20\degr40'55\farcs00$. The data were calibrated using the MIRIAD
package. Maps were made  with the visibility data weighted by the associated system
temperature and using natural weighting. Different taper functions were used to
improve the signal-to-noise ratio depending on the transition. Table
\ref{liniesobservades} lists characteristic parameters of channel as well as
continuum emission maps. The table gives the frequency of the transition, the
resulting synthesized beam, the channel resolution and the rms noise of the maps at
the channel resolution.

\section{Results and analysis}

\begin{figure}[fh]
\plotone{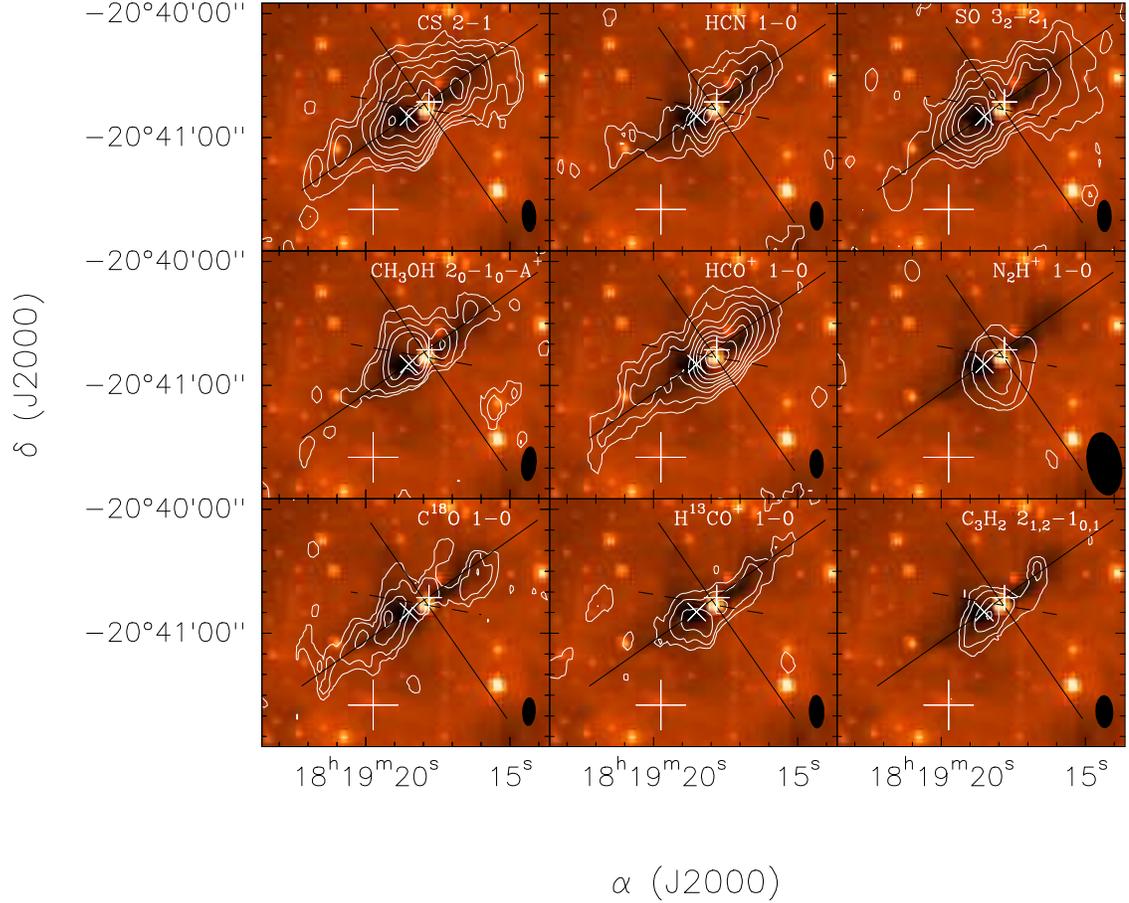}
\vspace{-2.0 cm}
\caption{
Superposition of the Spitzer image (8~$\mu$m; color scale) and the zero-order moment
(integrated emission; white contours) over the 9.5--14.09~\kms\ velocity range, of
the species detected  with BIMA in the HH 80N core. The Spitzer image shows the HH 80N core in
absorption against the background emission. The contour levels are 2, 4, 7,
10, 14, 18, 23, and 30 times the rms of each map: 0.14 (CS and SO), 0.11 (HCN, \methanol\ and \hcom), 0.1
(\htcom\ and \cthd), 0.17 (\cdo) and 0.40 (\ndhm) Jy~beam$^{-1}~$km~s$^{-1}$.  The
beam is shown in the bottom right corner of each panel. The large and small crosses
mark the HH 80N object and the continuum peak at 1 mm, respectively.  The tilted
cross marks the \ammonia\ peak (Girart et al. 1994). The solid lines represent the
minor (P.A. = 32\arcdeg) and the major (P.A. = 122\arcdeg) axes of the core. The
dashed line gives approximately the orientation and extension of the outflow 
detected by \dmu\ (P.A.
$\sim$~80\arcdeg). The
central bright source is likely associated with the 
source powering the molecular outflow.
\label{integrat}}
\end{figure}

Table~\ref{liniesobservades} lists the transitions (excluding hyperfine components)
of all detected molecules: \cdo, \hcom, \htcom, HCN, CS, SO, \ndhm, \cthd\ and
\methanol. The Table also lists the transitions of several molecules that were
undetected, among them SiO and H$_2$CO. The 1.4~mm continuum emission was detected
marginally with a flux density of $61\pm16$~mJy, peaking at
$\alpha(J2000)=18^\mathrm{h}19^\mathrm{m}17\fs81$ and
$\delta(J2000)=-20\arcdeg40'42\farcs70$, while no continuum emission was detected at
3.1~mm with an upper limit of $\sim$~12~mJy~beam$^{-1}$ (at the 3--$\sigma$ level).
Very Large Array 2 cm continuum observations with the D configuration (beamsize
$8.1''\,\times4.9''$) do not show emission towards the 1.4~mm continuum peak up to
0.27~mJy~beam$^{-1}$ (at the 3--$\sigma$ level) (Mart\'i et al.\ 1993). The
combination of 2~cm and 1.4~mm flux density yields $\alpha\geq$~2.0 where $\alpha$
is defined as $S_{\nu}\propto\nu^{\alpha}$, being $S_{\nu}$ the flux density at
frequency $\nu$. This lower limit value obtained for the spectral index is clearly
compatible with having thermal dust emission. Figure~\ref{integrat} shows the
integrated emission of the detected lines superimposed on the 8~$\mu$m Spitzer image
retrieved from the Spitzer archive. For reference, we define the bright central
source seen in the 8~$\mu$m Spitzer image as the core center
($\alpha(J2000)=18^\mathrm{h}19^\mathrm{m}17\fs81$ and
$\delta(J2000)=-20\arcdeg40'47\farcs74$). 


As seen in Fig.~\ref{integrat}, all the detected molecules (except for N$_2$H$^+$)
present an elongated structure with a position angle of $\sim120$\arcdeg\ (i.e. the
direction of the major axis) following approximately the same morphology revealed by
the dark lane seen in the 8~$\mu$m Spitzer image. Clearly, CS is the most extended
molecular tracer showing half-power angular dimensions of
60\arcsec$\,\times$~25\arcsec. SO, HCN and \hcom\ behave similarly to CS showing
approximately the same angular size, while \methanol, \cdo, \htcom\ and \cthd, present
smaller sizes, typically 50\arcsec$\,\times$~15\arcsec.  

\begin{figure}[fht]
\epsscale{0.8}
\begin{center}
\plotone{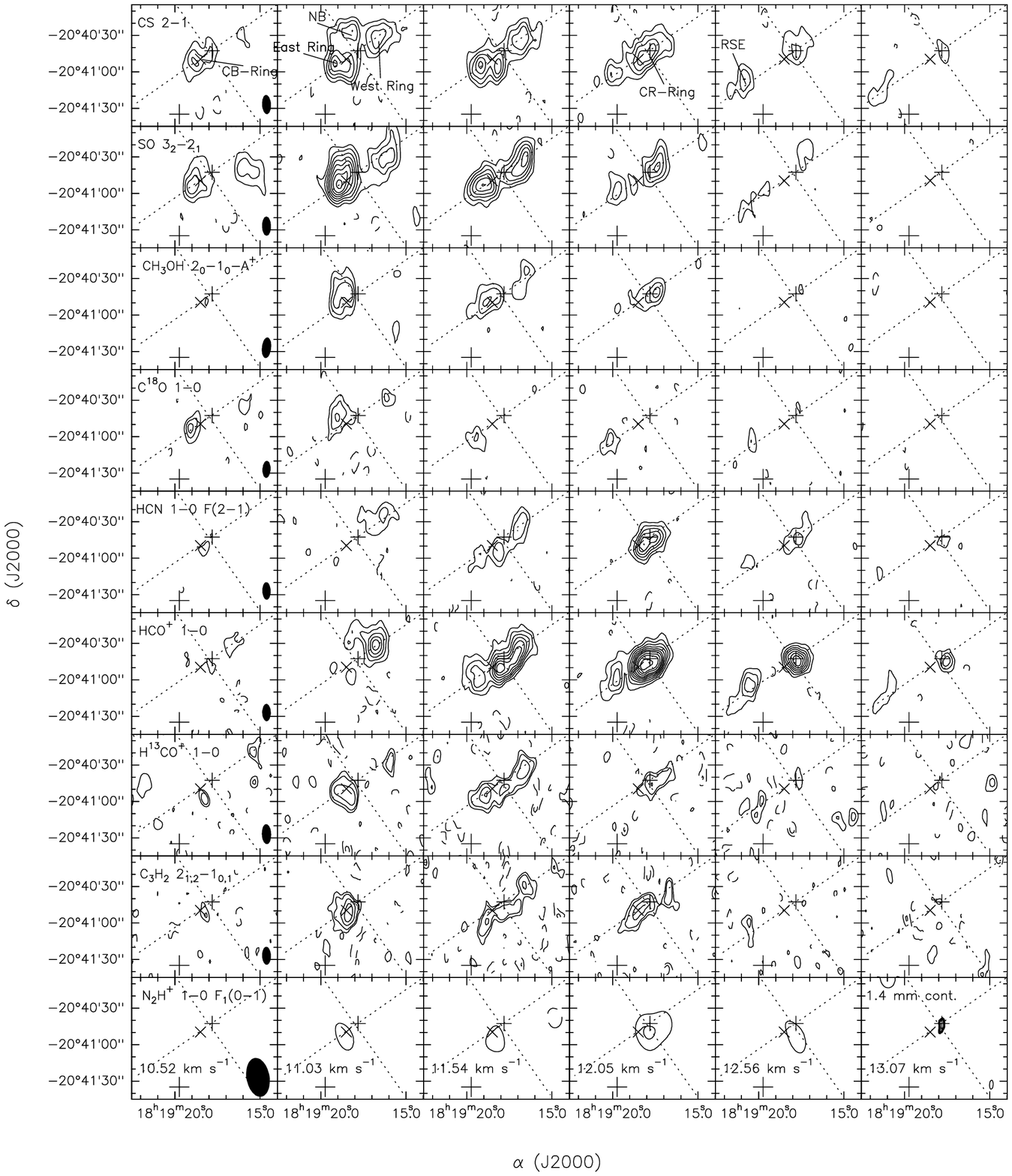}
\vspace{0 cm}
\figcaption{
Channel velocity maps of the species shown in Fig.~\ref{integrat} over the 10.5~\kms\ to
13.0~\kms\ velocity range, with a velocity resolution of $\sim0.5$~\kms. The level contours
are -3, 3, 5, 7, 9, 11, 13, 15, 18, and 21 times the rms noise level for each species. For
\htcom\ and \cthd\ the contours -2 and 2 times rms the noise are also shown. The labels
shown in the first row indicate the parts of the ring-like morphology (see \S~3), as well
as other structures. The thick solid contour shown in the bottom right panel represents 2
times the rms of the 1.4~mm continuum emission. The beam for each species (filled ellipse)
is shown in the bottom right corner of the panels of the first
column. The beam of the 1.4~mm continuum emission (open ellipse) is shown 
in the bottom right corner of the bottom right panel. The symbols are the same as in
Fig.~\ref{integrat}. \label{mapacanals} } 
\end{center} 
\end{figure}

Figure~\ref{mapacanals} shows the channel maps for the different species over the $\sim$~10.5
to $\sim$~13.1~\kms\ velocity range, where most of the emission is detected. The 1.4~mm
continuum emission is included in the bottom right panel of this figure. The emission of most of the
species has a clumpy morphology with different substructures along the core. As mentioned
above, \dmu\ analyzed the CS~(2--1) emission and found that the flattened structure is
consistent with a ring-like structure seen edge on. Thus, on basis of this result and for
clarity, we divided the core in the different parts of the ring analyzed by \dmu, plus some
other structures. 

The features indicated in the first row of Fig.~\ref{mapacanals} belonging to the ring are: (1) the
eastern side of the ring that appears as a clump seen at 11.0--11.5~\kms\ (hereafter we will
call this component East Ring); (2) the western side of the ring, that consists of a clump
visible at the same velocities as the East Ring (hereafter, West Ring); (3) the red-shifted
side of the ring that is seen at 12~\kms\ as a clump spatially coincident with the core center
(hereafter, CR-Ring); (4) in addition, we define the blue-shifted side of the ring at the
center of the core as the CB-Ring, seen in 10.52~\kms\ channel. The other features that do not
seem to belong to the ring are: (5) the eastern elongated structure found at red-shifted
velocities, 12.6--13.1~\kms, eastwards of the East Ring structure (hereafter, RSE, following
the nomenclature used in \dmu); (6) the northern blue-shifted clump seen only at 11~\kms\
(hereafter, NB). This nomenclature will be followed in the forthcoming analysis.  

The inspection of channel maps of Fig.~\ref{mapacanals} reveals an interesting
differentiation between molecular tracers. In addition, the kinematic complexity
becomes evident along the major axis of the core. The molecules can be grouped
according to the morphology of their emission:       

{\em (a)} CS, SO and \methanol. CS and SO are characterized by having the brightest
emission in the East Ring. They also trace other structures such as the West Ring, the
NB and the RSE. \methanol\ shows a similar morphology to CS and SO even though its
emission is more compact and it is weaker. In particular, the emission of \methanol\
in the RSE and CB-Ring is very marginal. 

{\em (b)} \hcom\ and HCN. The strongest emission of these species, especially
\hcom, arises in the CR-Ring where it shows a prominent clump. In addition,
these two species trace the RSE structure. However, the peculiarity of this
group of species is that they show little or no emission in the East Ring,
unlike the rest of species. The emission of \htcom, the optically thin
isotopologue of \hcom, appears more compact being only significant in the 11.0
to 12.0~\kms\ velocity channels. The \htcom\ emission is consistent with the
\hcom\ and HCN emission. However, the East Ring shown faintly by \hcom\ and
missed by HCN is clearly traced by \htcom. As the two species of this group have
a large dipole moment, the lack of emission in the most abundant \hcom\
isotopologue probably means that there is a foreground cold and low density
layer efficiently absorbing their emission (Girart et al. 2000).       

The rest of the species exhibit peculiar features and can not be included in any
group: the emission of \cthd\ is quite similar to that of \htcom\ and traces all the
structures except RSE and NB. On the other hand, \cdo\ only traces the East Ring and
the RSE structures. Finally, the \ndhm\ emission is more compact than the other
species not only due to the poor angular resolution, but to the poor signal-to-noise
ratio of the observations that would prevent us from detecting any extended
emission.

\subsection{Analysis of the morphology and kinematics}

\begin{figure}[fht]
\begin{center}
\vspace{-1.9 cm}
\plotone{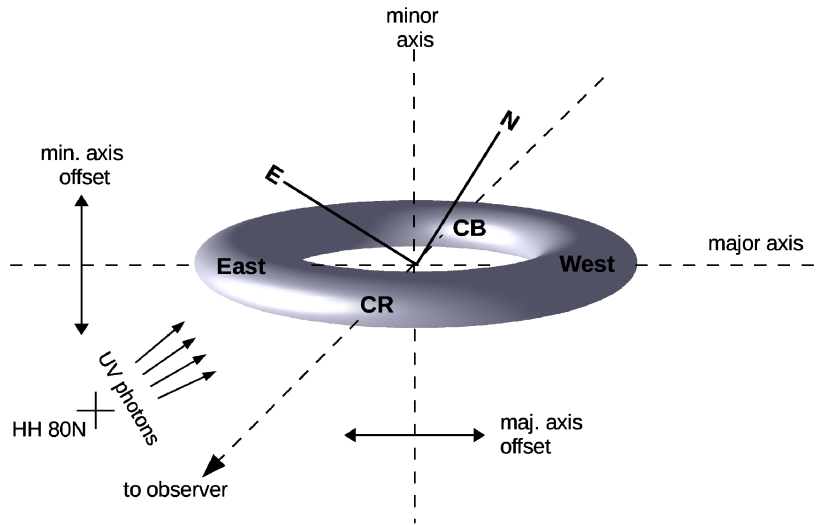}
\vspace{-1.0 cm}
\figcaption{
Sketch of the of the model geometry used in the analysis of the PV plots,  including a
schematic representation of the HH 80N scenario that we propose in order to account for the observed
chemistry in the ring components discussed in \S~4.2.2\label{ring}.  Note that 
this figure is not plotted to scale.}
\end{center} 
\end{figure}

To better study the kinematic signatures as well as the morphological differences
between the detected species, we performed a set of position-velocity (PV) plots along
the major (P.A.~$=122\arcdeg$) and minor (P.A.~$ = 32\arcdeg$) axis of the core.
Subsequently, we analyzed the properties of the emission for the different species by
modeling the emission as an optically thin and uniform contracting ring-like
structure. This model has been found to fit well the CS emission (\dmu). The model
consists, basically, in a spatially thin disk seen edge-on. A similar model was first
used by Ohashi et al.\ (1997) to study the emission of the contracting core associated
with IRAS $04368+2557$ in L1527. The parameters of our model are the inner and outer
radii of the ring, \rin\ and \rout, respectively, the infall velocity, \vinf, and the
intrinsic line width $\Delta$V. Our model does not account for rotation because \dmu\
found it to be negligible ($\le$~0.2~\kms), neither for dynamical infalling, for
simplicity. For clarity, we show in Fig.~\ref{ring} a 3D view of the model geometry.
This figure indicates the components of the ring defined in \S~3, as well as the
directions of minor and major axis. Our strategy is to obtain synthetic PV plots in
the minor and major axis of the edge-on ring-like structure, and to compare them with
the PV plots taken from the data. Although the ring model is assumed to be spatially
thin, we convolved the spatial axis of all the PV plots taken from the model with the
angular resolution of our observations. A similar procedure was performed in \dmu\ in
their attempt to study the CS emission. The authors obtained an inner and outer radii
of \rin$=23\arcsec$ and \rout$=35\arcsec$ ($\sim$~4$\times10^4$~AU and
$\sim$~6$\times10^4$~AU, respectively), an infall velocity of $\sim$~0.6~\kms, and a
line width of 0.8~\kms. In our analysis, we adopted the same model parameters as \dmu\
but leaving the inner and outer radii of the ring as free parameters. For consistency,
we first analyzed the CS~(2--1) data. Once the CS emission is modeled, it becomes a
reference for the rest of species. We found a \rout\ of 35\arcsec\
($\sim$~6$\times10^4$~AU) and a \rin\ of 15\arcsec\ ($\sim$~2.5$\times10^4$~AU) being
the latter value somewhat smaller than that obtained by \dmu. For other species, the
use of different values of \rin\ and/or \rout\ does not improve significantly the
appearance of the residual PV plots. Thus, in those cases we adopted the same \rin\
and \rout\ as for the CS.

\begin{figure}[fh]
\epsscale{0.7}
\plotone{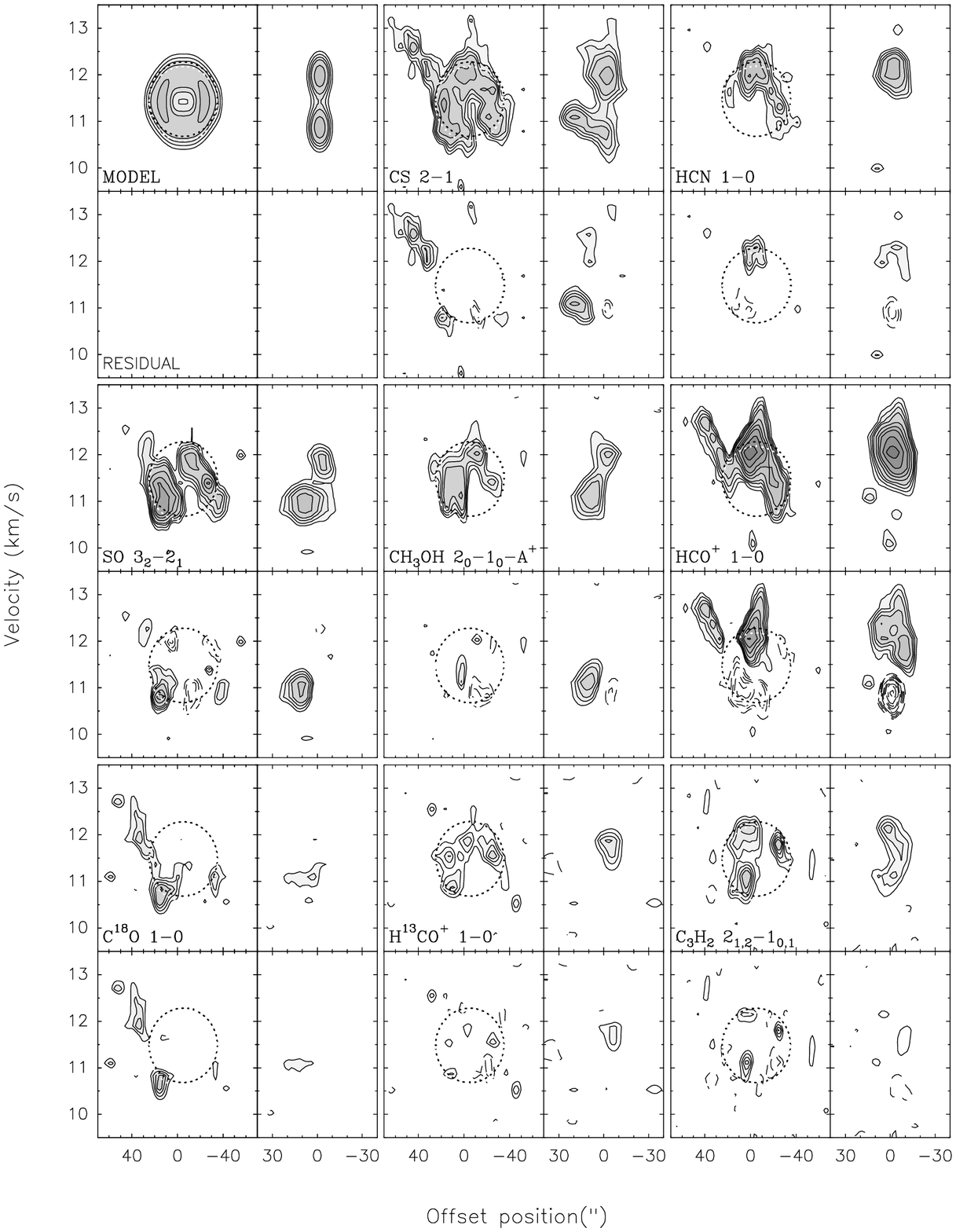}
\vspace{-0.3cm}
\figcaption{
Sets of PV plots for ({\it from left to right and from top to bottom}):
model, CS~(2--1),  HCN~(1--0), SO~(3$_2$--2$_1$), \methanol~(2$_0$--1$_0$--A$^+$),
\hcom~(1--0), \cdo~(1--0), \htcom~(1--0) and \cthd~(2$_{1,2}$--1$_{0,1}$). 
For each set, the PV plots of the BIMA data are shown at the top and the
residual PV plots resulting from subtracting the data and the model are shown at
the bottom. The panels on the left represent the PV plots along the major axis
while the panels on the right represent the PV plots along the minor axis of the
core. For the set of PV plots obtained from the model, the two panels at the top
show the best fit model (see text) for the major (left) and minor (right) axis
of the ring-like structure seen edge-on. The positive offsets of the PV
plots corresponds to South-East and to North-East
directions for the major and minor axis, respectively. The contour levels are -10,
-8, -6, -4, -3, 3, 4, 6, 8, 10, 12, 14, 16, and 20 times the rms noise for each
species. For \htcom\ and \cthd\ the contours -2 and 2 times the rms noise are also
shown. The dashed ellipse, drawn to highlight the best fit model for the major
axis, corresponds to a ring with a radius of 30\arcsec. 
\label{residu}
}
\end{figure}

Figure~\ref{residu} shows the PV plots along the major and minor axis of the emission, as
well as the residual PV plots after subtracting the best-fit model from the
data. As found by GEVWH, the model fits the CS emission reasonably well. However, the PV
plots for the other species reveal that they trace the contracting ring--like structure in
a non-uniform fashion.  The residual PV plots show some of the structures defined above as
important features of emission and \emph{absorption} for most species: 

\begin{itemize}

\item[-]\emph{RSE}. This eastern elongated structure, which is visible in CS, SO, \hcom,
\cdo\ and marginally in HCN, is clearly independent from the contracting ring as pointed out by
\dmu. The RSE exhibits a clear linear velocity gradient with increasing
velocity toward the east, i.e., as it gets closer to HH 80N.  

\item[-]\emph{NB}. It is clearly seen in the PV plots along the minor axis of
CS, SO and \methanol\ (and marginally \cdo). Due to its position (20\arcsec\ east
from the emission expected from the ring in the minor axis PV plot) this clump
is probably independent from the ring structure.   

\item[-]\emph{East Ring}. It appears as a spot seen in the residual PV plots along
the major axis of CS, SO and \cdo\ (and marginally \methanol) located at 11~\kms\
at the eastern part of the ring. Although its position indicates that the East
Ring seems to belong to the ring structure, the residual PV plots show that its
emission clearly exceeds the emission generated by the uniform ring model.

\item[-]\emph{West Ring}. Contrary to the other structures, the West Ring does not show
significant emission in the residual PV plots for almost all the species. Only CS, SO and
\cdo\ seem to show marginal emission in the western part of the ring structure. This implies
that for the majority of the species the molecular emission is consistent with the emission
expected for the ring model in this part of the ring.

\item[-]\emph{CR--Ring}. The CR-Ring seems to be physically part (at least partially)
of the ring structure but it exhibits stronger emission than that expected from the
ring model for HCN and, especially \hcom, as clearly seen in the residual PV plots.

\item[-]\emph{CB--Ring}. This structure is visible as negative contours in the residual PV
plots for the major and minor axis for most of the species. The CB-Ring component is best
traced by CS, but presents an important lack of emission for the rest of species, being
most evident for \hcom.

\end{itemize}

\subsection{Relative abundances}

\begin{figure}[f]
\epsscale{0.75}
\plotone{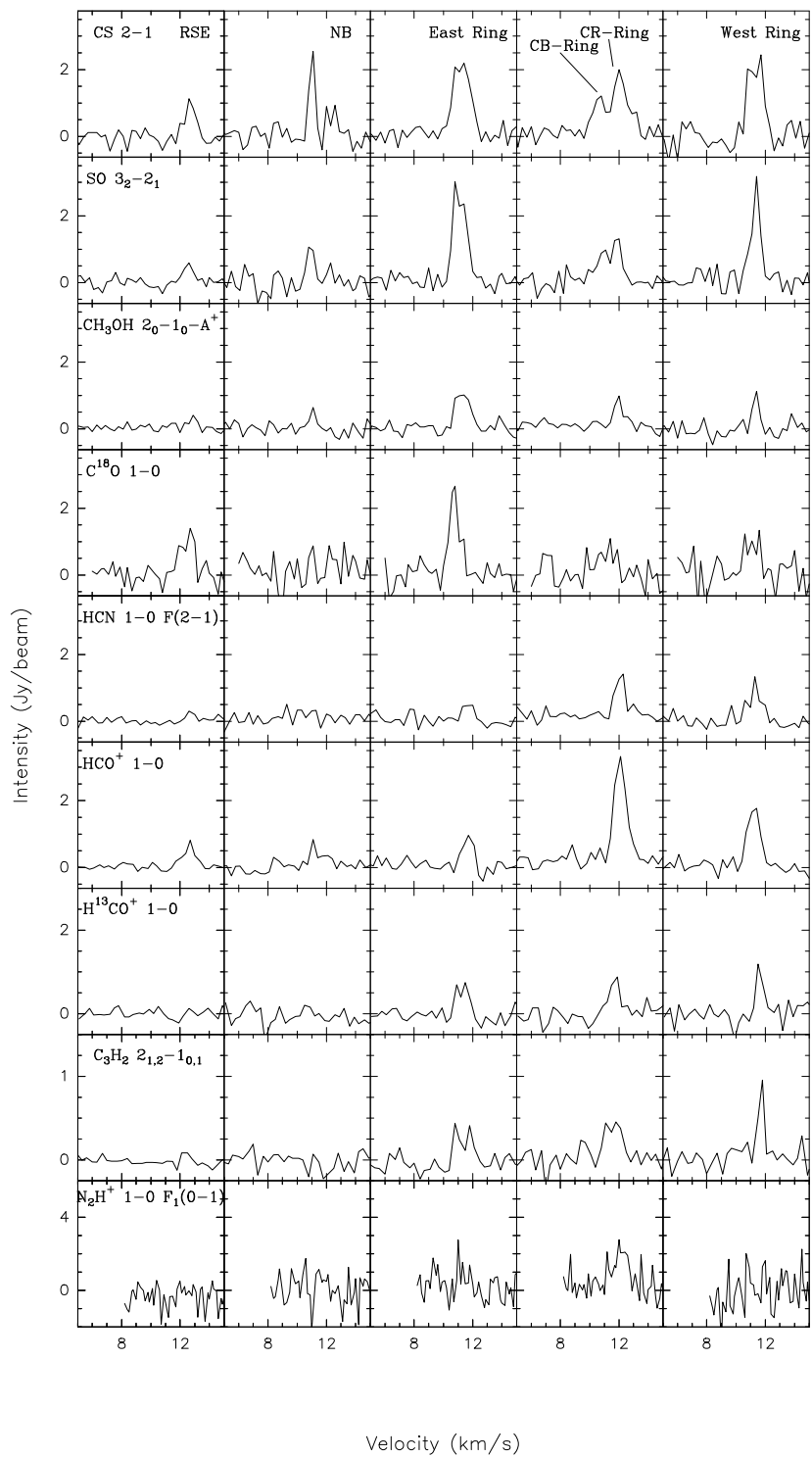}
\caption{Spectra of (\emph{from top to bottom}) CS~(2--1),
SO~($3_2$--$2_1$),  CH$_3$OH~(2$_0$--$1_0$)-A$^+$,
C$^{18}$O~(1--0), HCN~(1--0, F(2--1)),
HCO$^+$~(1--0),  H$^{13}$CO$^+$~(1--0),
C$_3$H$_2$~(2$_{1,2}$--$1_{0,1}$), and N$_2$H$^+$~(1--0, F$_1$(0--1)) towards (\emph{from left to right})
RSE, NB, East Ring, CB and CR-Ring, and West Ring components given in Table~\ref{positions}. 
For RSE, the spectra were obtained for a box of $\sim 15\arcsec \times 25\arcsec$.
 \label{espectres}}
\end{figure}

Table~\ref{positions} lists the positions that we consider most representative of the
structures defined previously. Figure~\ref{espectres} shows the spectra of the detected
transitions taken at the five selected positions shown in Table~\ref{positions}. Note
that some of the spectra taken at the center of the core present two velocity
components, separated by $\sim$~1~\kms. This result, especially evident for CS, is
perfectly compatible with the contracting ring-like morphology found previously for the
core and, in fact, these two velocity components appear to coincide with the CB--Ring
and CR--Ring components. Another general feature is that the spectra show narrow
linewidths, with a FWHM ranging from 0.6 to 1.4~\kms, which indicates that the gas in the
cores is not shocked (typical linewidths found in shocked regions are 6-7~\kms, e.g.
Bachiller et al.\ 1995).

From the spectra of Fig.~\ref{espectres} we derived the column densities in all
the components. To do so, we used RADEX, a non-LTE excitation and radiative
transfer code that can provide us constraints on  the gas density, kinetic
temperature and/or the column density (van der Tak et al.\ 2007). RADEX was used
adopting the temperature of 17~K, which is close to the rotational temperature
obtained from ammonia (Girart et al.\ 1994).  We used the \hcom~(3--2) and
\htcom~(3--2) lines observed by  Girart et al.\ (1998) in combination with the
\htcom~(1--0) (smoothed to the CSO angular resolution) of our dataset to
constrain the range of possible values for the average gas density. We
calculated line intensity ratios of these lines  assuming a comparable beam
dilution. Then, we ran the code to explore the densities that yield brightness
temperatures ratios compatible with our measured values (within a 20 per cent
range due to uncertainties). 

We found a possible range of 5 $\times$ 10$^4$~--~1.3
$\times$ 10$^5$~cm$^{-3}$ for the average gas density in the core. Once the
density range is constrained and the kinetic temperature fixed, using RADEX
again we were able to search for the column density range that yields the line
intensities measured in the spectra of Fig.~\ref{espectres}. For \hcom, the column
density was derived using the optically thin isotopologue \htcom\ and adopting a
carbon isotope abundance ratio, C$^{12}$/C$^{13}$, of 63 (Langer \& Penzias
1990; 1993). The results are presented in Table~\ref{column}.


In order to clarify the scenario that leads to the chemical properties observed in each
structure we computed the relative abundances with respect to CS. The reason for adopting
such a scale is that CS appears to be the most fiducial molecule since its chemistry seems
to be directly related to dense gas. In fact, it is the molecule that more regularly traces 
the contracting edge-on ring. In addition, as the lines involved in this analysis were taken
with the same configuration and beam sizes, then the missing flux effects are probably
canceled out. Another advantage of using the molecular abundances relative to CS is that
most of the observed transitions have a high critical density similar to CS~(2--1) and,
hence, they trace approximately the same gas (with the possible exception of \cdo). The
resulting ranges of fractional abundances with respect to CS corresponding to the density
range derived above are shown in Table~\ref{tabundancies}.

\begin{figure}[f]
\epsscale{0.65}
\plotone{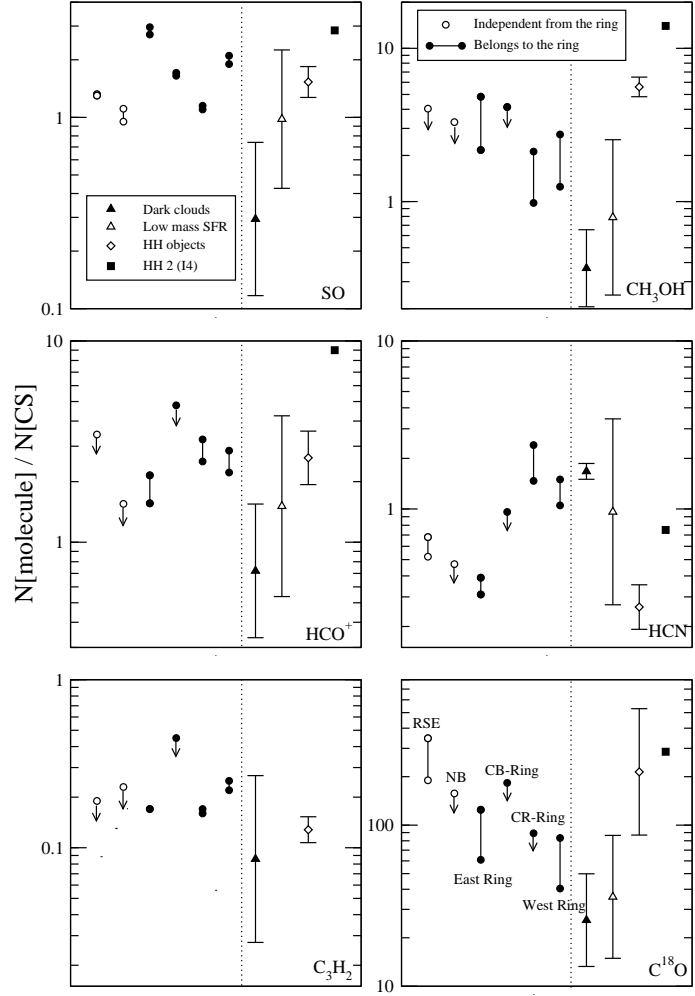}
\caption{
Relative molecular abundances (with respect to CS) for (\emph{from left to
right and from top to bottom}) SO, \methanol, \hcom, HCN, \cthd\ and \cdo. 
In each panel the circles represent the structures found for the HH 80N region: (\emph{from
left to right}) RSE, NB, East Ring, CB-Ring, CR-Ring and West Ring, and the bars between
circles represent the range of relative abundances derived from the density range found in \S~3.2.
The rest of the symbols and the bars represent the logarithmic median and
standard deviation of the relative molecular abundances for each
sample (\emph{from left to right}): dark clouds (i.e. without star formation
activity), low--mass star--forming cores, regions close to HH objects, 
and a position close to HH 2. The sample for the quiescent dark molecular clouds is: OMC-1N (Ungerechts et al.\ 1997),
TMC-1 (Pratap et al.\ 1997), L1498 and L1517B (Tafalla et al.\ 2006), L1544 and L1689B
(J\o rgensen et al.\ 2004), and B68 (di Francesco et al.\ 2002). The sample for the low mass star
forming molecular clouds is: IRAS $16293-2422$ (van Dishoeck et al.\ 1995), NGC 1333 IRAS
4A (Blake et al.\ 1995), IRAS $05338-0624$ (McMullin et al. 1994), Serpens S68 (McMullin
et al.\ 2000), several cores  from J\o rgensen et al. (2004) and Sch\"oier et al. (2002)
and \methanol\ data (Maret et al. 2005).  The sample of the regions close to HH objects
is: HH 2 (Girart et al.\ 2005), HH 1, several positions towards HH 7-11, and HH 34 (Viti
et al.\ 2006). The HH 2 data were taken in the SO$_2$ clump (see Girart et al.\ (2005)). 
\label{abundancies}}
\end{figure}

We also calculated the logarithmic median and the standard deviation of the relative
abundances for the same molecules detected in the HH 80N clump over a sample for three
types of environments related with dense gas found in the bibliography: molecular
clumps associated with HH objects, low-mass star-forming cores and dark molecular
clouds. Figure~\ref{abundancies} shows the comparison of the fractional abundances of
these three types of environments with the values of Table~\ref{tabundancies} (see figure 
caption for information on the sources of these samples).

The description of the chemical properties seen in Fig.~\ref{abundancies} led
us to classify the observed molecules depending on the variability of their
relative abundances between the different components of the core:  

{\em (a)} \cdo\ is the molecule that presents the major variability between
components. It exhibits a high relative abundance with respect to CS
towards RSE (between a factor of 2 to 5 with respect to the rest of components)
although it is undetected or has low relative abundances in the rest of the
components.

{\em (b)} SO, \methanol\ and \hcom\ have abundances higher than CS but presents
a moderate variability. The comparison between their relative abundances with respect
to CS in the different components of the core for the different species in this group
never exceeds a factor of 3. There is, however, a difference between the
molecules in this group: while SO and \methanol\ relative abundances are maximum at
the East Ring position, \hcom\ presents the highest abundances towards CR-Ring and
West Ring.

{\em (c)} \cthd\ and HCN exhibit a moderate variability and each molecule shows a distinct
behavior. Clearly, \cthd\ is poorly detected along the core. On the other hand, as seen in
\S~3, HCN has a similar morphology as \hcom. Since both molecules have comparable dipole
moments, the HCN emission may be considerably self-absorbed, similarly as \hcom. Therefore,
HCN relative abundances are likely underestimated in most components of the core.

\section{Discussion}

\subsection{Star formation activity within the core}

This work presents the results of a molecular survey carried out with
BIMA towards the star forming core ahead of HH 80N, for which the CS and CO data were already reported in
\dmu. From the CS data, \dmu\ detected an elongated dense core with 0.24 pc of
radius seen edge on. The properties of the CS emission suggested that it was
tracing a ring structure contracting with a striking supersonic infall
velocity (0.6~\kms). The central 'hole' of the ring is presumably the result of
molecular depletion due to the high central density that might be reached in
this core. From the CO data, \dmu\ found a bipolar outflow that is a signpost
of an ongoing star forming process in the core. The Spitzer 8~$\mu$m  and  the
BIMA 1.4~mm dust images show a bright source located at the center of the
bipolar CO outflow. This strongly suggests the presence of an embedded
protostellar object within the core, which powers the bipolar outflow.

The CS emission was used in \dmu\ to derive the mass of the ring structure. Their
obtained value, which is 12~$M_\odot$ assuming $X$(CS)~=~$2\times$10$^{-9}$
(10~$M_\odot$ belonging to the main core, and 2~$M_\odot$ to RSE), is somewhat
lower than the value obtained from \ammonia\ (20~$M_\odot$, Girart et al.\ 1994).
The 1.4~mm continuum emission is barely resolved, which implies that it arises from
a region  with a size of $\sim$~10$^4$~AU in diameter. The mass of the dust can be
estimated using the formula
\begin{equation}
M_d=\frac{S_\mathrm{1.4} D^2 c^2}{2kT_\mathrm{d}\nu^2k_\mathrm{1.4}} 
\end{equation}
where $S_\mathrm{1.4}$ is the continuum flux measured at 1.4~mm, $D$ is the distance to the source,
$T_\mathrm{d}$ is the dust temperature, and $k_\mathrm{1.4}$ is the mass opacity coefficient at
1.4~mm.  In our calculations we assume a $k_\mathrm{1.4}$ of 0.83~cm$^2$~gr$^{-1}$ obtained from
$k_\mathrm{1.3}$ for a density of $n_\mathrm{H_2}$~$\simeq$~10$^6$~cm$^{-3}$ (Col.~2 in Table 2 of
Ossenkopf \& Henning\ 1994) and assuming a dust opacity index of 2, and a temperature range of
20--40~K (Jennings et al.\ 1987), which is among the temperatures expected from classical
theoretical models of protostellar envelopes (e.g. Adams \& Shu\ 1985). Note that this expression
uses the Rayleigh-Jeans approximation, valid at millimeter wavelengths. For our values the mass of
the dust is 0.019-0.036~$M_\odot$. Adopting a gas-to-dust ratio of 100, the total mass of the
envelope traced by the dust emission is 1.9-3.6~$M_\odot$


Bontemps et al.\ (1996) found a relationship between the outflow momentum flux,
the envelope mass and the bolometric luminosity. \dmu\ found that the envelope
mass was almost an order of magnitude higher than the values predicted by Bontemps
et al., for the measured CO momentum flux. However, Bontemps et al. derived the
mass of the envelope for the sources in their sample by integrating the 1.3~mm
emission over a region corresponding to $\sim$~0.05~pc (10$^4$~AU), whereas \dmu\
used the mass derived from the CS emission, which traces a region with a radius of
0.24~pc. The derived mass from the 1.4~mm dust corresponds to a region of size
approximately similar to that used in Bontemps et al. Thus, if we use the total
mass of the envelope derived from the 1.4~mm flux, then the relationship between
the envelope mass and the CO outflow momentum flux fits very well.

The study of the circumstellar mass surrounding the YSOs at several radii is a
useful tool to track their evolution (Andr\'e \& Montmerle\ 1994). Class 0 YSOs
are best interpreted as very young protostars whose dense circumstellar envelope
still contains an important reservoir of mass to be accreted. Therefore the fact
that the bulk of the mass of the HH 80N core is found at large radii from its
central embedded YSO suggests that this object should be at the beginning of the
main accretion phase and, hence, it is another signpost of being an extremely
young YSO found probably in the Class 0 stage (e.g. Girart et al. 2009). 

\subsection{Chemical properties}

The analysis of the other species of the molecular survey confirms that their
morphology roughly follows the ring-like morphology seen in CS, contracting
with a supersonic infall velocity ($\sim$~0.6~\kms), though this is a first-order
result: a more accurate analysis reveals that most of the molecular species,
although tracing approximately the same region as CS, show important local
differences. This deviation reflects a possible combination effect of excitation,
optical depth and molecular abundance, being the latter the most plausible
candidate to explain most of the observed differences. 

In this section, we attempt a simple discussion on the possible origin of the
observed differentiation by means of a qualitative comparison with the chemistry
observed in several environments associated with the dense gas. We investigate whether
the chemistry observed in the HH 80N core is the result of a strong
central depletion and of photochemical effects caused by the UV radiation
coming from the HH 80N object, as pointed out by \dmu. 


\subsubsection{The ring-like morphology and the central depletion}

As suggested by \dmu, the ring-like structure is likely due to the freeze--out of
the molecules onto the dust grains at the inner regions of the core, where high
densities are expected. In addition, there is a young protostellar object embedded
in the core. Thus, the conditions of the central part of the core should resemble
those found in protostellar cores. In this section we analyze the
inner structure of the core by comparing it qualitatively with other similar
regions.

When modeling the radial variation of molecular abundances in a circumstellar
envelope one has to take into account the effects of the central embedded YSO,
heating the inner part of the core, which causes partial sublimation of the icy
dust mantles (e.g. IRAS 16293-2422: Ceccarelli et al.\ 2000; Sch\"oier et al\
2002;  NGC1333: Maret et al.\ 2002, 2004; J\o rgensen et al.\ 2005;
Bottinelli et al.\ 2007). However, we do not see any signpost of a warm component
(temperatures of several tens of K) in the HH 80N core, which could be due to either
the low luminosity or the youth of the YSO, as well as the distance of the
region (in other words, the expected warm region is likely highly beam diluted).


Outside the undetected warm inner region, the HH 80N core exhibits a large region where
molecules seem to be strongly frozen in the icy mantles, except possibly for \ammonia\
(Girart et al.\ 1994) and \ndhm, molecules that seem to exhibit a more compact emission.
Unfortunately, from the analysis of the PV plots we are not able to determine whether other
species deplete at different radii from the center of the core than that of CS.  In any
case, the depletion radius of the HH 80N core ($\sim$~2.5$\times10^4$~AU) is significantly
larger (by 3--4 times) than those obtained for prestellar cores by Tafalla et al.\ (2006). 
Because the densest gas traced by molecular species is expected to be found at the inner
radius of the ring, the lower value of the possible range of average densities for the HH
80N core found in \S~3.2 (5~$\times$~10$^4$~cm$^{-3}$) can be taken as a lower limit for
the density at the depletion radius (i.e. depletion density).  As the Class 0 object
forming inside the HH 80N core is very young, the core probably still retains the initial
conditions for star formation. Hence, there might be some physical connection with the
prestellar cores, specially with those that are found on the verge of dynamical collapse.
The CS is depleted at densities of the order of a few 10$^4$~cm$^{-3}$ in
prestellar cores (Tafalla et al.\ 2002; Pagani et al.\ 2005), slightly below the lower
limit for the depletion density in the HH 80N core. The large depletion radius and
densities found for the HH 80N core suggests high central densities. This could be due to a
more evolved stage of the HH 80N core with respect to prestellar cores, favored by its
rapid collapse.  



Surrounding the `frozen' region, the HH 80N core has a molecular shell or ring-like
structure. Since a minimum gas density is required to observe a given molecular
transition, the detection of some specific transitions of our dataset can provide an
estimation of the lower possible densities of this shell (i.e. at outer radius 6
$\times$ 10$^4$~AU). In particular, the detection of \hcom~(1--0) line implies densities
higher than $2.4 \times 10^3$~cm$^{-3}$ (for which this transition is easily
detectable, Evans et al.\ 1999). In summary, we conclude that the densities traced by
the molecular emission associated with the ring-like structure are above $\sim2 \times
10^3$~cm$^{-3}$, reaching values likely higher than 5~$\times$~10$^4$~cm$^{-3}$ at the inner
radius.



\subsubsection{The components of the ring-like structure}

In Figure~\ref{ring} we present the most likely scenario as seen by an hypothetical
observer. In the following we analyze the enhancement of some species on
the basis of this scenario and in the frame of the UV irradiation. Our main consideration
is the effect of the distance of each component with HH 80N. The fact that HH 80N may not
lie in the same plane of sky as the core does not affect qualitatively our analysis. 
This picture implies higher photochemical activity in the East Ring and moderate
or low photochemical effects for the CR-Ring, West Ring and CB-Ring. In particular,
Table~\ref{column} shows that the column densities of CS, \methanol, \cdo\ and SO follow
fairy well this trend: most of them are higher in the East Ring than in the West Ring. 

\begin{figure}[f]
\epsscale{1.}
\plotone{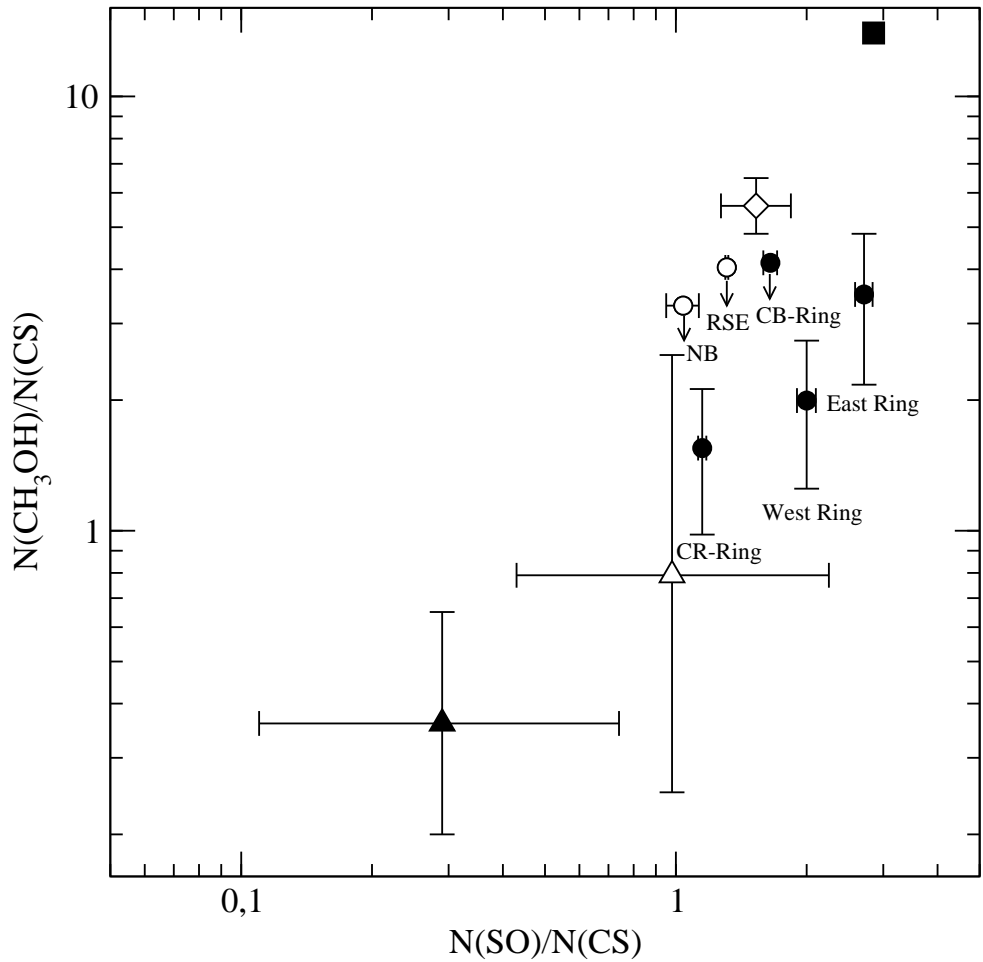}
\caption{
\methanol\ vs. SO relative molecular abundances (with respect to CS). The circles represent the
structures found for the HH 80N region and the bars associated with the circles represent the range of relative abundances for
the density range found in \S~3.2. The rest of the symbols and bars are the same as in
Fig.~\ref{abundancies}.
\label{so_ch3oh}}
\end{figure}

One of the most evident results of Fig.~\ref{abundancies} is that the enhancement of SO in
the East Ring component (the closest in projection to HH 80N) resembles that obtained for
the clump ahead of the HH~2 jet (Girart et al.\ 2005) and determined to be due to chemical
effects induced by an enhanced radiation field. This trend is also observed for \methanol\
and \cdo. Figure~\ref{so_ch3oh} also shows that SO and \methanol\ are  higher than the
typical range of values of low mass star forming cores. The caveat to this scenario is
\hcom, which presents similar column densities in all the components of the ring (with the
exception of CB-Ring) instead of being greater in the East Ring and CR-Ring, as one would
expect if the radiation field impinging on the ring were higher than the interstellar
radiation field. This result indicates that other parameters, apart from the effect of the
distance to HH 80N, also play a role on the chemistry of the core. Clearly, a detailed
modeling, which is however beyond the scope of this paper, is required in order to
characterize the chemistry of the core ahead of HH 80N.



Molinari et al.\ (2001), using a simple model, derived a value of 670~G$_0$ for the FUV
field emanating from the recombination region in HH 80N. From this value, only the fraction
of photons that impinges to the HH 80N core over the total has to be taken in account.
Taking 0.2~pc as the distance between the core and the HH object, and assuming that the
core is seen edge on from HH 80N, we obtain a dilution factor of 1/16, which yields
$\sim$~40~G$_0$ for the side of the core facing HH 80N (the East Ring).  Interestingly,
this value is within the range of possible values for the radiation field expected to
impinge the clump ahead of HH 2 (Viti et al.\ 2003).  However, note that the dilution
factor derived above is based on a projected distance. Therefore, its value as well as the
estimated FUV are upper limits.

On the other hand, based on the range of average density values estimated in \S~3.2, the
derived column density through the ring seen from the edge of the core is $N$(H$_2)$ = $2.5
- 6.5\times 10^{22}$~cm$^{-2}$. Then, the derived the visual extinction, $A_\mathrm{v}$,
from the relationship $N$(H$_2) \simeq 10^{21} A_\mathrm{v}$ (Spitzer\ 1978) is between
25--65~mag, much greater than the values where the photochemical effects are attenuated
(e.g. Viti \& Williams\ 1999).  Although HH 80N does not lie exactly on the same plane of
the ring, these high extinction values are only approximate and indicate that the UV
photons hardly penetrate deep inside the core. Therefore, any photochemical effects
that may affect the East Ring are likely to happen on the "surface" of the ring. 

The decrease of photochemical activity in the parts of the ring further away than the East
Ring is possibly a combined effect of geometrical dilution and the presence of a low density
cloud component. In fact, the distance from HH 80N to the western parts of the ring is a
factor of 2--3 greater than that to the illuminated side ($\sim$~0.2~pc), and much larger than the
typical distances between other HH objects and their respective associated clumps
($\sim$~0.05~pc, Viti et al.\ 2006). 






\subsubsection{Other molecular clumps in the region}

The RSE and NB structures were poorly detected in the species of our survey.
Both structures appear to be spatially independent from the ring structure in
addition of being smaller. Therefore, these components may present different
chemistry simply because their basic physical structure might be different from
the physical structure of the ring. There is, however, an important difference
between RSE and NB: while \cdo\ exhibits a high relative abundance with respect
to CS in RSE reaching values typical of those found in the HH 2 region (I4
position), it is undetected in NB. In addition,  \hcom\ is not enhanced with
respect standard molecular cloud values in NB, implying that the UV radiation 
does not play a significant role in the chemistry of this structure.



\section{Conclusions}

We have carried out an observational study with the BIMA array of the star
forming dense core ahead of HH 80N in order to characterize its morphological
and chemical properties. We present maps of the different molecular species
detected with BIMA (CS, \hcom, \htcom, HCN, \methanol, SO, \cdo, and \cthd) as
well as 1.4~mm continuum dust emission. These maps reveal a complex morphology
and kinematics for this core. In an attempt to understand this complexity, we
obtained PV plots for all the species along the major and minor axis of the core
and modeled them in a similar procedure as in \dmu. Subsequently, we derived the
column density and the relative abundances with respect to CS in some selected
positions along the core. Our main conclusions can be summarized as follows: 

\begin{enumerate}

\item[1.] The 1.4~mm dust emission was detected marginally in a position close
to the Spitzer 8~$\mu$m source. This source is located at the center of the CO
bipolar outflow found by \dmu\ and, therefore, it is likely its powering
source. The 1.4~mm dust emission traces the inner and denser region
($\la10^4$~AU) of the dense core around the protostar. The mass of the dense
core (1.9--3.6~$M_\odot$) at scales up to $\simeq10^4$~AU and the momentum flux
of the CO bipolar outflow correlate well with what is expected for very young
Class 0 sources (Bontemps et al.\ 1996).

\item[2.] The integrated emission for most of the molecular tracers reveals a
clumpy and elongated morphology with a size of $\sim$~60\arcsec$\,
\times$~25\arcsec\  (0.5~pc~$\times$~0.2~pc) and a position angle of
$\sim120\arcdeg$. Three main structures can be distinguished: the central core
around the protostar and two smaller clumps (RSE and NB). As previously observed
in CS by \dmu, the molecular emission associated with the central core arises
from the outer shell of the dense core. This molecular shell has a ring-like
morphology seen edge-on, which is well-traced by CS and partially traced by the
other molecular species.  The inner and outer radii of the ring are
$\sim2.5\times10^4$~AU and $\sim6.0\times10^4$~AU, respectively, and the estimated
averaged density ranges between $5\times10^4$~cm$^{-3}$ and
$1.3\times10^5$~cm$^{-3}$.  The ring structure has been divided in four regions,
following the CS clumpy morphology: the East Ring, Central Red-shifted Ring
(CR--ring), the Central Blue-shifted Ring (CB--ring) and the West Ring. The
kinematics traced by the different molecules are similar to that found for CS by
\dmu, which is indicative of contraction with a supersonic infall velocity of
0.6~\kms.

\item[3.] The relative abundances with respect to CS derived in different positions
were compared with those obtained for a sample of dark clouds, low-mass star-forming
cores and other molecular clumps ahead of HH objects.  This comparison suggests that
the relative abundances of SO, \cdo\ and, to a lesser extent \methanol, are partially
compatible with the high level of UV irradiation generated by HH 80N. In particular,
the section of the molecular ring facing HH 80N (East Ring), as well as RSE,
seem to be the regions more exposed to the UV radiation, whereas in the other
sections of the ring and NB the photochemical effects seem to be less important. The
molecular ring has a visual extinction, $A_\mathrm{v}$, between 25 and 65~mag seen
from the edge. These values are much higher than the maximum extinction at which the
UV radiation can effectively penetrate the ring and release the molecular species
from the dust mantles triggering the photochemistry (Viti et al.\ 2003). 
Inside the
molecular ring, at scales of $\la~2.5\times10^4$~AU, where extremely high densities
are found, the dense core exhibits a ``frozen'' region, where the observed molecules
are significantly depleted onto icy dust mantles.

\end{enumerate}

\acknowledgments

We thank the anonymous referee for his valuable comments, which helped to improve the
paper. JMG, RE, MBT and JMM are supported by MEC grant AYA2005-05823-C03 (co-funded
with FEDER funds). JMG and JMM are partially supported by AGAUR grant 2005 SGR 00489.
SV acknowledges financial support from an individual STFC (ex-PPARC) Advanced Fellowship.

\clearpage

\begin{table}[ht]
\scriptsize
\caption{Characteristic values of BIMA channel maps. \label{liniesobservades}}
\begin{tabular}{lr@{.}lcr@{~$\times$~}lr@{.}lr@{.}lr@{.}lr@{.}l}
\tableline
\tableline
Transition &  \multicolumn{2}{c}{$\nu$} & & \multicolumn{2}{c}{Beam\tablenotemark{a}} & \multicolumn{2}{c}{PA} &\multicolumn{2}{c}{rms} & \multicolumn{2}{c}{$\Delta
v_{ch}$} & \multicolumn{2}{c}{Peak emission}\tablenotemark{b} \\
          &    \multicolumn{2}{c}{GHz}  & Configuration &\multicolumn{2}{c}{arcsec$\times$arcsec} & \multicolumn{2}{c}{degrees} &
	  \multicolumn{2}{c}{(Jy~beam$^{-1}$)}&\multicolumn{2}{c}{(km~s$^{-1}$)}&  \multicolumn{2}{c}{(Jy~beam$^{-1}$)}\\              
\tableline
HCO$^+$~(1--0)           & 89&1885  &C&  14.1&6.9  & 1&2 &0&13&0&33&3&65\\
HCN~(1--0)                & 88&6319  &C& 14.2&6.7 & 1&7 &0&13&0&33&1&55\\
C$_3$H$_2$~(2$_{1,2}$--$1_{0,1}$)& 85&3389&C& 14.7&7.0 &1&8 &0&13&0&34&0&87\\
CS$~(2$--1)                 & 97&9810  &C& 15.6&7.1 &3&0 &0&18&0&30&2&36\\
C$^{18}$O~(1--0)          & 109&7822 &C& 13.9&6.5 &$-2$&9&0&29&0&27&2&77\\
SO~($2_3$--$1_2$)             & 109&2522 &C& 14.2&6.6&$-2$&0&0&24&0&40&$\leq$0&72\\
SO~($2_1$--$1_1$)             & 86&7543  &C& 18.1&7.8 &0&6&0&15&0&58&$\leq$0&45\\
SO~($3_2$--$2_1$)             & 99&2999  &C& 15.6 & 7.2 &0&7 &0&21&0&29&3&05\\
SO~($5_5$--$4_4$)             & 215&2207 &C&  7.8&3.8 &$-2$&4 &0&93&0&41&$\leq$2&79\\
CH$_3$OH~(2$_n$--$1_n$)       & 96&7414\tablenotemark{c}&C& 16.9&7.5 &$-6$&0&0&18&0&30&1&33\\
N$_2$H$^+$~(1--0)         & 93&1738\tablenotemark{d}&D& 32.1&18.8 &10&0&0&73&0&16&6&01\\
H$^{13}$CO$^+$~(1--0)     & 86&7543  &C& 16.2&7.6 &2&0 &0&16&0&34&0&99\\
C$_2$S~($6_7$--$5_6$)         & 86&1841  &C& 17.2&7.9 &1&1 &0&12&0&58&$\leq$0&36\\
SiO~(2--1)                & 86&8461  &C& 18.1&7.9 &$-0$&8&0&12&0&32&$\leq$0&36\\
HCOOH~(4$_{0,4}$--$3_{0,3}$)    & 89&5792  &C& 17.6&7.8 &$-0$&9&0&16&0&55&$\leq$0&48\\
H$_2$CS~(3$_{0,3}$--$2_{0,2}$)  & 103&0404 &C& 14.5&6.2 &2&0 &0&15&0&57&$\leq$0&45\\
H$_2$CO~(3$_{0,3}$--$2_{0,2}$)  & 218&2222 &C& 7.7&3.7 &$-9$&5&0&94&0&40&$\leq$2&82\\
Continuum~(1.4 mm)                          &  217&0	&C& 7.3&3.0&$-4$&5&0&016& \multicolumn{2}{l}{---} &0&061\\
Continuum~(3.1 mm)                          &  99&5	&C&15.5&6.8&$-6$&7&0&004& \multicolumn{2}{l}{---} &$\leq$0&012\\
\tableline
\end{tabular}
\vspace{1cm}
\tablenotetext{a}{
A taper function of 4\arcsec\ was used in all the channel maps
 except for C$^{18}$O(1--0) (2.0\arcsec), SO($5_5$--$4_4$) (2.5\arcsec),
 SiO(2--1) (2.0\arcsec), H$_2$CS(3$_{03}$--$2_{02}$)  (2.0\arcsec),
 H$_2$CO(3$_{03}$--$2_{02}$) (2.5\arcsec) and continuum (1.4~mm) (1.0\arcsec) maps.}
\tablenotetext{b}{~Upper limits are 3 times the rms.}
\tablenotetext{c}{~Frequency for the 2$_0$--$1_0$--A line. The
2$_{-1}$--$1_{-1}$--E line was also detected.}
\tablenotetext{d}{~Frequency for the F$_1$~F=2$_3$--1$_2$ hyperfine
line. All the hyperfine 1--0 components were detected.}
\end{table}

\begin{table}[ht]
\caption{Representative positions. \label{positions}}
\begin{tabular}{lccl}
\tableline
\tableline
     &\multicolumn{2}{c}{Position}&  \\
\cline{2-3}
Structure         & $\alpha$~(J2000) & $\delta$~(J2000) &Map counterpart\tablenotemark{a} \\
\tableline
RSE      & $18^\mathrm{h}19^\mathrm{m}21\fs10$
         & $-20\arcdeg41'15\farcs0$
         & CS$~(2$--1) $>$~12~\kms\ eastern elongation \\
NB       & $18^\mathrm{h}19^\mathrm{m}18\fs10$    
         & $-20\arcdeg40'26\farcs9$ 
	 & CS$~(2$--1) 11~\kms~northern peak\\
East Ring& $18^\mathrm{h}19^\mathrm{m}19\fs03$    
         & $-20\arcdeg40'53\farcs0$ 
	 & SO~($3_2$--$2_1$) 11~\kms~eastern peak  \\  
CB-Ring \& CR-Ring       & $18^\mathrm{h}19^\mathrm{m}17\fs81$    
         & $-20\arcdeg40'47\farcs7$ 
	 & HCO$^+$~(1--0) 12~\kms~center peak   \\ 
West Ring& $18^\mathrm{h}19^\mathrm{m}16\fs57$    
         & $-20\arcdeg40'31\farcs5$ 
	 & CS$~(2$--1) 11~\kms~western peak   \\ 

\tableline  
\end{tabular}
\tablenotetext{a}{Molecule used to better highlight the structure.}
\end{table}

\begin{table}[ht]
\tiny
\caption{Column densities\tablenotemark{a} (in units of 10$^{13}$~cm$^{-2}$)\label{column}}
\begin{tabular}{l|r@{.}lr@{.}l|r@{.}lr@{.}l|r@{.}lr@{.}l|r@{.}lr@{.}l|r@{.}lr@{.}l|r@{.}lr@{.}l|rr}
\tableline
\tableline
Molecule  & \multicolumn{4}{c}{CS} & \multicolumn{4}{c}{SO}  & \multicolumn{4}{c}{CH$_3$OH\tablenotemark{b}} & \multicolumn{4}{c}{HCN} &
\multicolumn{4}{c}{HCO$^+$\tablenotemark{d}} & \multicolumn{4}{c}{C$_3$H$_2$} & \multicolumn{2}{c}{C$^{18}$O}  \\                             
\tableline
& \multicolumn{2}{c}{$N_\mathrm{m}$} & \multicolumn{2}{c}{$N_\mathrm{M}$}& \multicolumn{2}{c}{$N_\mathrm{m}$} & \multicolumn{2}{c}{$N_\mathrm{M}$}& \multicolumn{2}{c}{$N_\mathrm{m}$}
& \multicolumn{2}{c}{$N_\mathrm{M}$} & \multicolumn{2}{c}{$N_\mathrm{m}$} & \multicolumn{2}{c}{$N_\mathrm{M}$} & \multicolumn{2}{c}{$N_\mathrm{m}$} & \multicolumn{2}{c}{$N_\mathrm{M}$}
&\multicolumn{2}{c}{$N_\mathrm{m}$} & \multicolumn{2}{c}{$N_\mathrm{M}$}& $N_\mathrm{m}$ & $N_\mathrm{M}$\\
\tableline
RSE &        0&9  & 0&5  &  1&2  &  0&7  &  $\leq$~1&8\tablenotemark{c}  &  $\leq$~2&1\tablenotemark{c}  &  0&6 &   0&3  & $\leq$~2&5\tablenotemark{c}  &
$\leq$~1&8\tablenotemark{c}  &  $\leq$~0&2\tablenotemark{c}  &  $\leq$~0&1\tablenotemark{c}  &  171 & 180 \\
NB  &        2&2  &  1&0  &  2&1  &  1&1  &  $\leq$~3&0\tablenotemark{c}  &  $\leq$~3&4\tablenotemark{c} &   $\leq$~1&0\tablenotemark{c} &  $\leq$~0&5\tablenotemark{c}   & 
$\leq$~2&2\tablenotemark{c}  & $\leq$~1&6\tablenotemark{c}  &  $\leq$~0&5\tablenotemark{c}  & $\leq$~0&2\tablenotemark{c}  &  $\leq$~154\tablenotemark{c}&
$\leq$~162\tablenotemark{c} \\
East Ring &  4&6  &  2&4  &  13&6 &  6&4  &  10&0 & 11&4  &  1&8 &  0&7   &  7&2 &  5&1  &   0&8  &  0&4  &  280 &  293\\
CB-Ring  &   1&2  &  0&7  &  2&1   & 1&2  &  $\leq$~2&6\tablenotemark{c}  & $\leq$~2&9\tablenotemark{c}  &  $\leq$~1&2\tablenotemark{c}  &  $\leq$~0&5\tablenotemark{c}   &
$\leq$~4&8\tablenotemark{c}  &   $\leq$~3&4\tablenotemark{c}   &  $\leq$~0&5\tablenotemark{c}  &   $\leq$~0&3\tablenotemark{c}   &  $\leq$~123\tablenotemark{c}   & 
$\leq$~130\tablenotemark{c}  \\
CR-Ring  &   2&8  &  1&5  &  3&1   & 1&7  &  2&7  & 3&1  &  6&6  &  2&2   & 6&9 &  4&7   & 0&5  &  0&2   &  $\leq$~123\tablenotemark{c}   & $\leq$~130\tablenotemark{c}  \\
West Ring &  4&1  &  2&1  &  8&7   & 4&0  &  5&1  & 5&8  &  6&2  &  2&2   & 9&1 &  6&0   & 1&0  &  0&5   &  166  & 175 \\
\end{tabular}
\tablenotetext{a}{~$N_\mathrm{m}$ and $N_\mathrm{M}$ are the column densities derived for the lower ($5 \times 10^4$~cm$^{-3}$) and upper ($1.3 \times 10^5$~cm$^{-3}$)
values of the density range derived from
the RADEX analysis (see \S~3.2)}
\tablenotetext{b}{~Including CH$_3$OH-A and CH$_3$OH-E.}
\tablenotetext{c}{~Upper limits are 3 times the rms noise.}
\tablenotetext{d}{~Derived from H$^{13}$CO$^+$.}
\end{table}

\begin{table}[ht]
\footnotesize
\caption{Relative abundances with respect to CS\tablenotemark{a}\label{tabundancies}}
\begin{tabular}{l|r@{.}lr@{.}l|r@{.}lr@{.}l|r@{.}lr@{.}l|r@{.}lr@{.}l|r@{.}lr@{.}l|rr}
\tableline
Molecule  & \multicolumn{4}{c}{SO}  & \multicolumn{4}{c}{CH$_3$OH\tablenotemark{b}} & \multicolumn{4}{c}{HCN} &
\multicolumn{4}{c}{HCO$^+$} & \multicolumn{4}{c}{C$_3$H$_2$} & \multicolumn{2}{c}{C$^{18}$O}  \\                             
\tableline
& \multicolumn{2}{c}{$X_\mathrm{m}$} & \multicolumn{2}{c}{$X_\mathrm{M}$}& \multicolumn{2}{c}{$X_\mathrm{m}$} & 
\multicolumn{2}{c}{$X_\mathrm{M}$} &\multicolumn{2}{c}{ $X_\mathrm{m}$ }& \multicolumn{2}{c}{$X_\mathrm{M}$} &\multicolumn{2}{c}{ $X_\mathrm{m}$} 
& \multicolumn{2}{c}{$X_\mathrm{M}$}&\multicolumn{2}{c}{ $X_\mathrm{m}$ }&\multicolumn{2}{c}{$X_\mathrm{M}$}& $X_\mathrm{m}$ & $X_\mathrm{M}$\\
\tableline 
RSE &        1&3 & 1&3 & $\leq$~2&0\tablenotemark{c} & $\leq$~4&0\tablenotemark{c} 
& 0&7 & 0&5 &  $\leq$~2&7\tablenotemark{c} & $\leq$~3&4\tablenotemark{c} 
& $\leq$~0&2\tablenotemark{c} &  $\leq$~0&2\tablenotemark{c} &    190 &  346 \\
NB  &        1&0 & 1&1 & $\leq$~1&4\tablenotemark{c} & $\leq$~3&3\tablenotemark{c}
 & $\leq$~0&5\tablenotemark{c} &$\leq$~0&4\tablenotemark{c} &  $\leq$~1&0\tablenotemark{c} & $\leq$~1&6\tablenotemark{c} 
 & $\leq$~0&2\tablenotemark{c} &  $\leq$~0&2\tablenotemark{c} &    $\leq$~72\tablenotemark{c}  &  $\leq$~157\tablenotemark{c} \\
East Ring &  3&0 & 2&7 & 2&2 & 4&8 & 0&4 & 0&3 &  1&6 & 2&2 & 0&2 &  0&2 &    61  &  125 \\
CB-Ring  &   1&7 & 1&7 & $\leq$~2&1\tablenotemark{c} & $\leq$~4&1\tablenotemark{c} & $\leq$~1&0\tablenotemark{c} & $\leq$~0&7\tablenotemark{c} &  
$\leq$~4&0\tablenotemark{c} & $\leq$~4&8\tablenotemark{c} & $\leq$~0&5\tablenotemark{c} &  $\leq$~0&4\tablenotemark{c} &   $\leq$~103\tablenotemark{c}  &  $\leq$~183\tablenotemark{c} \\
CR-Ring  &   1&1 & 1&2 & 1&0 & 2&1 & 2&4 & 1&5 &  2&5 & 3&2 & 0&2 &  0&2 &   
 $\leq$~45\tablenotemark{c}  &  $\leq$~89\tablenotemark{c} \\
West Ring &  2&1 & 1&9 & 1&3 & 2&7 & 1&5 & 1&0 &  2&2 & 2&8 & 0&3 &  0&2 &    40  &   83 \\
\end{tabular}
\tablenotetext{a}{~$X_\mathrm{m}$ and $X_\mathrm{M}$ are the abundance values derived for the lower ($5 \times 10^4$~cm$^{-3}$) and upper (1.3$ \times 10^5$~cm$^{-3}$)
limits of the density range derived from the RADEX analysis (see \S~3.2).}
\tablenotetext{b}{~Including CH$_3$OH-A and CH$_3$OH-E.}
\tablenotetext{c}{~3--sigma upper limits.}
\end{table}

\clearpage




\end{document}